\documentclass[]{aastex631}

\usepackage{graphicx}	
\usepackage{amsmath}	
\usepackage{amssymb}	
\usepackage{float}
\usepackage{graphicx}
\usepackage{subfigure}
\usepackage{multirow}
\usepackage{enumerate}
\usepackage[figuresright]{rotating}
\def\degree{${}^{\circ}$}

\submitjournal{ApJ}

\shorttitle{QUASI-PERIODIC VARIATIONS OF CORONAL MASS EJECTIONS}
\shortauthors{Xia Li et al.}
\graphicspath{{./}{figures/}}

\begin{document}

\title{Quasi-periodic Variations of Coronal Mass Ejections with Different Angular Widths}

\correspondingauthor{Hui Deng,Feng Wang, Linhua Deng and Ying Mei}
\email{denghui@gzhu.edu.cn, fengwang@gzhu.edu.cn, lhdeng@ynao.ac.cn, meiying@gzhu.edu.cn}

\author{Xia Li}
\affiliation{Center For Astrophysics, Guangzhou University,
Guangzhou,Guangdong, 510006, P.R. China}
\affiliation{Great Bay Center, National Astronomical Data Center,
Guangzhou, Guangdong, 510006, P.R. China}

\author{Hui Deng}
\affiliation{Center For Astrophysics, Guangzhou University, 
Guangzhou,Guangdong, 510006, P.R. China}
\affiliation{Great Bay Center, National Astronomical Data Center,
Guangzhou, Guangdong, 510006, P.R. China}

\author{Feng Wang}
\affiliation{Center For Astrophysics, Guangzhou University,
Guangzhou,Guangdong, 510006, P.R. China}
\affiliation{Great Bay Center, National Astronomical Data Center,
Guangzhou, Guangdong, 510006, P.R. China}

\author{Linhua Deng}
\affiliation{Yunnan Observatories, Chinese Academy of Sciences,
Kunming, Yunnan, 650216, P.R. China}

\author{Ying Mei}
\affiliation{Center For Astrophysics, Guangzhou University,
Guangzhou,Guangdong, 510006, P.R. China}
\affiliation{Great Bay Center, National Astronomical Data Center,
Guangzhou, Guangdong, 510006, P.R. China}



\begin{abstract}
Coronal mass ejections (CMEs) are energetic expulsions of organized magnetic features from the Sun. The study of CME quasi-periodicity helps establish a possible relationship between CMEs, solar flares, and geomagnetic disturbances. We used the angular width of CMEs as a criterion for classifying the CMEs in the study. Based on 25 years of observational data, we systematically analyzed the quasi-periodic variations corresponding to the CME occurrence rate of different angular widths in the northern and southern hemispheres, using frequency and time-frequency analysis methods. There are various periods for CMEs of different angular widths: 9 months, 1.7 years, and 3.3-4.3 years. Compared with previous studies based on the occurrence rate of CMEs, we obtained the same periods of 1.2(±0.01) months, 3.1(±0.04) months, $\approx$6.1(±0.4) months, 1.2(±0.1) years, and 2.4(±0.4) years. We also found additional periods of all CMEs that appear only in one hemisphere or during a specific solar cycle. For example, 7.1(±0.2) months and 4.1(±0.2) years in the northern hemisphere, 1(±0.004) months, 5.9(±0.2) months, 1(±0.1) years, 1.4(±0.1) years, and 2.4(±0.4) years in the southern hemisphere, 6.1(±0.4) months in solar cycle 23 (SC23) and 6.1(±0.4) months, 1.2(±0.1) years, and 3.7(±0.2) years in solar cycle 24 (SC24). The analysis shows that quasi-periodic variations of the CMEs are a link among oscillations in coronal magnetic activity, solar flare eruptions, and interplanetary space.

\end{abstract}

\keywords{Sun: coronal mass ejections (CMEs) - Sun: solar-terrestrial relations - Sun: magnetic fields}


\section{Introduction} 
\label{1}

Coronal mass ejections (CMEs) are one of the intensest manifestations of solar magnetic activity that significantly affect the interplanetary environment and space weather \citep{Chen2011Coronal,2019ApJS2449W}. For example, CMEs create heliospheric disturbances, driving shocks, and accelerating electrons and protons evidenced by radio bursts and solar energetic particles (SEPs), which are one of the major causes of geomagnetic storms on Earth. Determining the temporal evolution characteristics of CMEs, especially their possible periodic patterns, is valuable for establishing the unique correlations among CMEs, intense solar flares, and geomagnetic disturbances \citep{2012Coronal,2013ApJ76537F,2013ApJ76343C}.

With the operation of the Large Angle Spectrometric Coronagraph (LASCO) \citep{1995The}, many high-quality observational CME data have been obtained. Scientists gradually began to use these data to carry out studies of CME periodicity. Table \ref{tab:1} lists some information about their research, including the quasi-periods of the CMEs, the catalogs they used, the time range of the observational data, and the analysis methods. 

The prime parameter for such studies is the occurrence rate of the CMEs \citep{lou2003periodicities,2008Short,2014Different,2014Wavelet,2018Periodic,lamy2019coronal}. Another two parameters often used are the mass rate \citep{2010Comprehensive,2018Periodic} and the speed of the CMEs \citep{kilcik2019comparison,kilcik2020temporal}. The most popular catalog used in these studies is Coordinated Data Analysis Workshops catalog (CDAW) \citep{lou2003periodicities,2008Short,2014Different,2014Wavelet,kilcik2019comparison,kilcik2020temporal}. \citet{2018Periodic} and \citet{lamy2019coronal} utilized the ARTEMIS catalog for their research. 

The periodic analysis methods commonly applied are wavelet analysis, Fourier transforms, maximum entropy, and Lomb-normalized periodogram. The quasi-periodicity of CMEs obtained from the study mainly includes the quasi-biennial oscillations (QBOs) and the Rieger-type periodicity. QBOs have periodicities of around 2 years, a type of mid-term quasi-periodicity considered to be related to the solar dynamic process and the emergence of the solar magnetic flux \citep{2001ARep51012O,2002SoSyR..36..507K,2014SSRv86359B,deng2019phase}. The periods ranging from 130 to 220 days are called the Rieger-type periods that are thought to be related to the periodic emergence of magnetic flux from the deep solar interior \citep{ballester2002near}. The formation of a CME typically involves a perturbation in the low corona, usually in the form of a solar flare \citep{hudson2001observing}. Although there is no one-to-one relationship between CME occurrence rate and solar flares, a causal link between these two eruptive events may be pronounced such that any flare accompanying a CME is part of an underlying magnetic process \citep{2012Coronal}. And there are also some studies on the relationship of the time and the period between CMEs and flares \citep{kilcik2019comparison,kilcik2020temporal}. 
In addition, some studies further discussed the periods of CMEs in the northern and southern hemispheres, solar cycle 23 (SC23) and solar cycle 24 (SC24), respectively \citep{2018Periodic, lamy2019coronal}. 

However, few previous works studied the period of CMEs with different angular widths. The angular widths of CMEs are measured as the position angle extent in the sky plane. \citet{2017ApJ84979Z} reported on the importance of the angular width of a CME in determining whether the corresponding interplanetary CME and the preceding shock will reach Earth. \cite{gopalswamy2002interacting} suggested that partial halo CMEs ( 120\degree\textless angular width\textless 360\degree) have an important role in producing large SEPs. While the average width of the radio-poor CMEs is much smaller than that of the radio-rich CMEs \citep{1029/2001JA000234}. And \citet{kahler2001coronal} proposed that narrow CMEs (angular width$\leq$20\degree) are often associated with impulsive (not gradual) SEPs. Therefore, the quasi-periodicity of CMEs with different angular widths can reveal whether CMEs with different geomagnetic effectiveness might have different periods. 

In this study, we use the latest CDAW catalog to study the quasi-periodicity of CMEs with different angular widths based on the CME occurrence rate. In Section \ref{2} we describe the data and the two methods used for periodic analysis. The results are presented in Section \ref{3}. Section \ref{4} presents discussions. We conclude the study in Section \ref{5}.

\begin{table}
\centering
\caption{The main previous studies on the quasi-periodicity of CMEs published over the past two decades.}
\label{tab:1}
\begin{tabular}{lllll}
\hline
\multicolumn{5}{c}{Occurrence rate}                                                                                                                                          \\
\hline
Paper                         & Periods                                                & Catalog                   & Time                        & Method                    \\
\hline
                              &                                                        &                           &                             & Fourier                   \\
\multirow{-2}{*}{\citet{lou2003periodicities}}       & \multirow{-2}{*}{1.2 Mo,3.4 Mo,6.1 Mo,11.3 Mo}         & \multirow{-2}{*}{CDAW}    & \multirow{-2}{*}{1999-2003} & Wavelet                   \\
                              & 3.1 Mo,6.3 Mo                                          &                           &                             & Maximum entropy           \\
\multirow{-2}{*}{\citet{2008Short}}      & 1.1 Yr                                                 & \multirow{-2}{*}{CDAW}    & \multirow{-2}{*}{1996-2006} & Wavelet                   \\
                              &                                                        &                           &                             & Fourier                   \\
\multirow{-2}{*}{\citet{2014Different}} & \multirow{-2}{*}{5.1 Mo,6.3 Mo}                        & \multirow{-2}{*}{CDAW}    & \multirow{-2}{*}{1999-2012} & Wavelet                   \\
\citet{2014Wavelet}                      & 4.2-8.4 Mo                                             & CDAW                      & 2000-2012                   & Wavelet                   \\
                              &                                                        &                           &                             & Fourier                   \\
\multirow{-2}{*}{\citet{2018Periodic}}   & \multirow{-2}{*}{3.2 Mo,1 Yr,2.4 Yr}                   & \multirow{-2}{*}{ARTEMIS} & \multirow{-2}{*}{1996-2016} & Wavelet   etc.            \\
                              & {\color[HTML]{333333} }                                &                           &                             & Fourier                   \\
\multirow{-2}{*}{\citet{lamy2019coronal}}        & \multirow{-2}{*}{{\color[HTML]{333333} 3.2 Mo,2.4 Yr}} & \multirow{-2}{*}{ARTEMIS} & \multirow{-2}{*}{1996-2018} & Wavelet                   \\
\hline
\multicolumn{5}{c}{Occurrence rate + N/S}                                                                                                                                    \\
\hline
                              & N:3.1 Mo,5.9 Mo,1.2 Yr,2.5 Yr,3 Yr                     &                           &                             & Fourier                   \\
\multirow{-2}{*}{\citet{2018Periodic}}   & S:6.5 Mo,1.7 Yr,1.9 Yr                                 & \multirow{-2}{*}{ARTEMIS} & \multirow{-2}{*}{1996-2016} & Wavelet   etc.            \\
                              & N:5.9 Mo,1.2 Yr                                        &                           &                             & Fourier                   \\
\multirow{-2}{*}{\citet{lamy2019coronal}}        & S:6.4 Mo,1.9 Yr                                        & \multirow{-2}{*}{ARTEMIS} & \multirow{-2}{*}{1996-2018} & Wavelet                   \\
\hline
\multicolumn{5}{c}{Occurrence rate + SC23/SC24}                                                                                                                              \\
\hline
                              & SC23:3.2 Mo,6.8 Mo,1.1 Yr,2.4 Yr                       &                           &                             &                           \\
\multirow{-2}{*}{\citet{2018Periodic}}   & SC24:                                                  & \multirow{-2}{*}{ARTEMIS} & \multirow{-2}{*}{1996-2016} & \multirow{-2}{*}{Wavelet} \\
                              & SC23:3.2 Mo,6.8 Mo,1.1 Yr,2.4 Yr                       &                           &                             &                           \\
\multirow{-2}{*}{\citet{lamy2019coronal}}        & SC24:                                                  & \multirow{-2}{*}{ARTEMIS} & \multirow{-2}{*}{1996-2018} & \multirow{-2}{*}{Wavelet} \\
\hline
\multicolumn{5}{c}{Mass rate}                                                                                                                                                \\
\hline
\citet{2010Comprehensive}                   & 5.9 Mo                                                 & CDAW                      & 1996-2009                   & Lomb-normalized           \\
                              & {\color[HTML]{333333} }                                &                           &                             & Fourier                   \\
\multirow{-2}{*}{\citet{2018Periodic}}   & \multirow{-2}{*}{{\color[HTML]{333333} 1.7 Yr}}        & \multirow{-2}{*}{ARTEMIS} & \multirow{-2}{*}{1996-2016} & Wavelet   etc.            \\
                              &                                                        &                           &                             & Fourier                   \\
\multirow{-2}{*}{\citet{lamy2019coronal}}        & \multirow{-2}{*}{1.7 Yr}                               & \multirow{-2}{*}{ARTEMIS} & \multirow{-2}{*}{1996-2018} & Wavelet                   \\
\hline
\multicolumn{5}{c}{Mass rate + N/S}                                                                                                                                          \\
\hline
                              & N:1.6 Yr,3 Yr                                          &                           &                             & Fourier                   \\
\multirow{-2}{*}{\citet{2018Periodic}}   & S:6.1-6.2 Mo,1.7-1.8 Yr,2 Yr                           & \multirow{-2}{*}{ARTEMIS} & \multirow{-2}{*}{1996-2016} & Wavelet   etc.            \\
                              & N:                                                     &                           &                             & Fourier                   \\
\multirow{-2}{*}{\citet{lamy2019coronal}}        & S:2.2 Mo,6.2 Mo,1.7-1.9 Yr                             & \multirow{-2}{*}{ARTEMIS} & \multirow{-2}{*}{1996-2018} & Wavelet                   \\
\hline
\multicolumn{5}{c}{Mass rate + SC23/SC24}                                                                                                                                    \\
\hline
                              & SC23:2.3 Mo,5 Mo,6.9 Mo,1 Yr,1.7 Yr                    &                           &                             &                           \\
\multirow{-2}{*}{\citet{2018Periodic}}   & SC24:1.8 Yr                                            & \multirow{-2}{*}{ARTEMIS} & \multirow{-2}{*}{1996-2016} & \multirow{-2}{*}{Wavelet} \\
                              & SC23:2.3 Mo,5 Mo,6.9 Mo,1 Yr,1.7 Yr                    &                           &                             &                           \\
\multirow{-2}{*}{\citet{lamy2019coronal}}        & SC24:1.8 Yr                                            & \multirow{-2}{*}{ARTEMIS} & \multirow{-2}{*}{1996-2018} & \multirow{-2}{*}{Wavelet} \\
\hline
\multicolumn{5}{c}{Maximum CME speed index (MCMESI)}                                                                                                                         \\
\hline
                              & 1 Mo,1.1-1.2 Mo,1.4 Mo,2 Mo                            &                           &                             & Multi Taper               \\
                              & 2.5 Mo,3.1 Mo,5.1 Mo,5.9-7.1 Mo                        &                           &                             & Wavelet                   \\
\multirow{-3}{*}{\citet{kilcik2019comparison}}    & 1.1 Yr,1.3 Yr                                          & \multirow{-3}{*}{CDAW}    & \multirow{-3}{*}{1996-2018} &                           \\
                              & 0.9-1.2 Mo,2.5 Mo,10-10.4 Mo                           &                           &                             & Multi Taper               \\
\multirow{-2}{*}{\citet{kilcik2020temporal}}    & 0.97-1.1 Yr,1.5 Yr,2.8-3.2 Yr                          & \multirow{-2}{*}{CDAW}    & \multirow{-2}{*}{1999-2018} & Wavelet                   \\
\hline
\multicolumn{5}{l}{Note. Month=Mo, Year=Yr, N=Northern hemisphere, S=Southern hemisphere.}                                                                                  
\end{tabular}
\end{table}

\section{Data Preparation and Period Analysis Methods}
\label{2}

\def\degree{${}^{\circ}$}

\subsection{Data Source}
\label{2.1}

We selected the CDAW catalog \citep{Yashiro2004A,gopalswamy2006coronal} for our research. There are five catalogs from LASCO observations. Among them, only CDAW is manually maintained and hence is commonly used for research by scientists (see Table \ref{tab:1}). CDAW provides the information of each CME, including
time, date, angular width, central position angle (CPA), and so on. The daily occurrence rate of the CMEs is shown in Figure \ref{fig:Figure1}. There are 31,310 CMEs in total. The data cover the period from January 1, 1996, to August 31, 2021, while two gaps exist: June 26, 1998, to October 9, 1998, and December 21, 1998, to February 2, 1999, a total of 150 days. We supplemented the gaps with a linear interpolation method.

\subsection{Classification of CMEs by Angular Width}
\label{2.2}

\begin{figure}
\begin{center}
	\includegraphics[width=\textwidth]{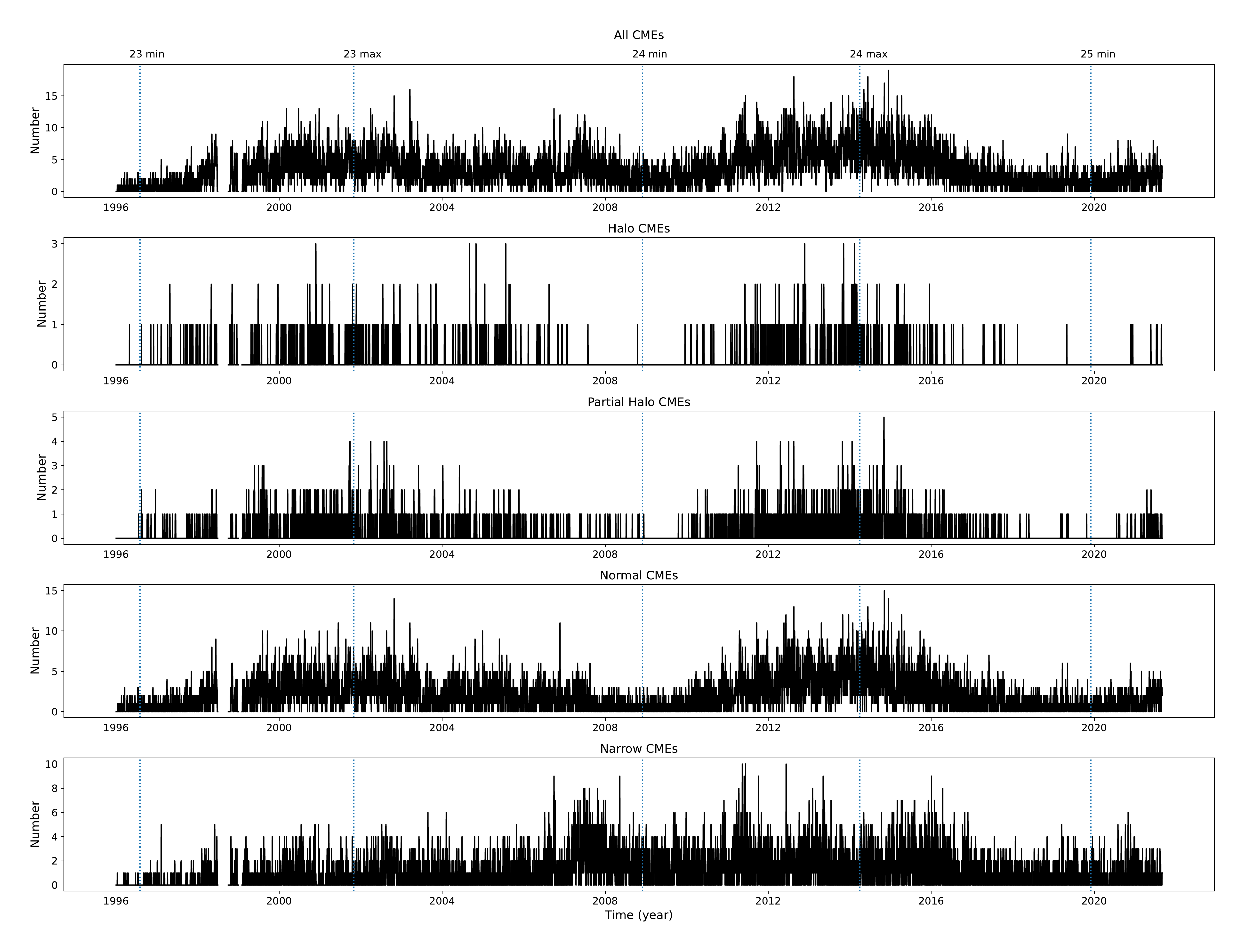}
\end{center}
    \caption{The daily occurrence rate of CMEs (CDAW catalog). The panels from top to bottom show all/halo/partial halo/normal/narrow CMEs, respectively.}
    \label{fig:Figure1}
\end{figure}

To investigate the quasi-periods of the CMEs with different angular widths, we classified them into four types: halo CMEs (angular width=360\degree), partial halo CMEs (120\degree\textless angular width\textless 360\degree), normal CMEs (20\degree\textless angular width$\leq$120\degree), and narrow CMEs (angular width$\leq$20\degree), following the classification method proposed by \citet{Yashiro2004A}. The number of each type of CMEs is listed in Table \ref{tab:2}, respectively. From Figure \ref{fig:Figure1} and Table \ref{tab:2}, it can be found that most of the CMEs belong to the normal (accounting for 59.4\%) and narrow (accounting for 32.1\%) CMEs. These two types of CMEs account for 91.5\% of the total number.   

\subsection{Classification of CMEs by Angular Width and Hemisphere}
\label{2.3}

\begin{table}
\centering
\renewcommand\tabcolsep{6.5pt} 
\caption{The number and percentages of different types.}
\label{tab:2}
\begin{tabular}{lccccc}
\hline
                                         & All       & Halo   & Partial halo  & Normal    & Narrow    \\
\hline
Whole Time (January 1996 to August 2021) & 31310(100\%)  & 734(2.3\%) & 1932(6.2\%)       & 18607(59.4\%) & 10037(32.1\%) \\
\hline
Northern hemisphere                      & 15306(48.9\%) & -          & 979(3.1\%)        & 9320(29.8\%)  & 5007(16.0\%)  \\
Southern hemisphere                      & 14956(47.8\%) & -          & 926(3.0\%)        & 9076(29.0\%)  & 4954(15.8\%)  \\
\hline
\end{tabular}
\end{table}

\begin{figure}
\begin{center}
	\includegraphics[width=\textwidth]{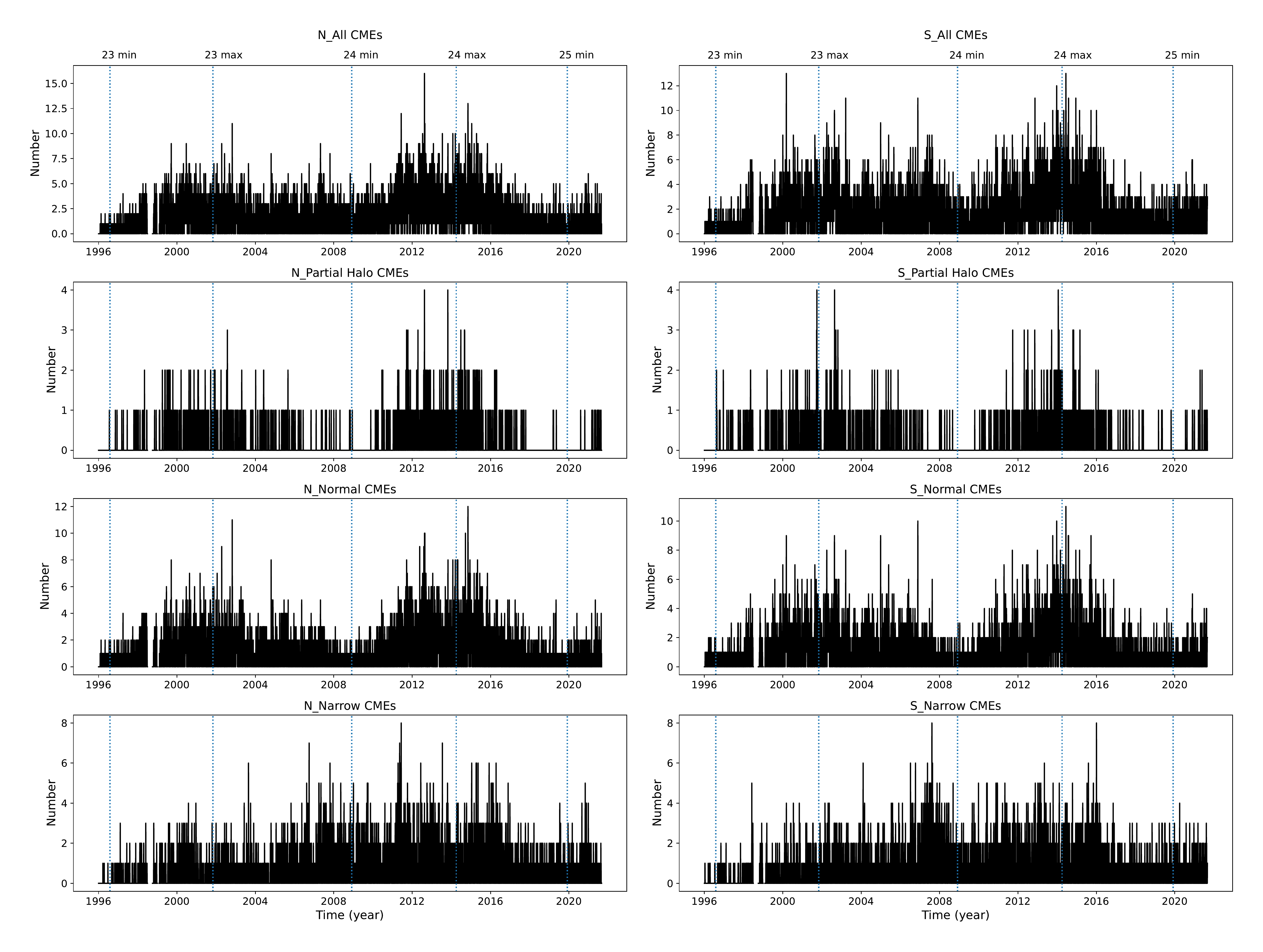}
\end{center}
    \caption{The daily occurrence rate of CMEs (CDAW catalog) in the northern and southern hemispheres. The left column is for the northern hemispheres. The right column is for the southern hemispheres. From top to bottom panels show all/partial halo/normal/narrow CMEs, respectively.}
    \label{fig:Figure2} 
\end{figure}

According to the CPA provided by the CDAW catalog, we classified the CMEs into two groups: the CMEs in the northern hemisphere and those in the southern hemisphere. The halo CMEs are excluded for this purpose. It needs to note that 314 events ($\approx$1\%) are centered at the equator, which are classified into neither northern nor southern hemisphere. Figure \ref{fig:Figure2} presents the time series of the daily occurrence rate of the CMEs in the northern and southern hemispheres. Table \ref{tab:2} lists the number of the three types of CMEs in each hemisphere.

\subsection{Period Analysis Methods}
\label{2.4}

Frequency analysis (Fourier spectra) and time-frequency analysis (wavelet spectra) \citep{farge1992wavelet} were applied to investigate the quasi-periodicity of CMEs with different angular widths. In spectral analysis, as pointed out by \cite{pardo2015comparison} and \citet{2018Periodic}, the peaks in the spectra may be real or noise resulting from errors in the detection of events and the measurements of physical quantities. Therefore, it is necessary to have reasonable estimates of the power spectrum. Meanwhile, assessing the reliability of statistical significance of those estimates is equally essential. Each methodology has its parametric statistical test. Therefore, we also estimated statistical significance for the Fourier spectral analysis and wavelet spectral analysis, respectively. 

1. We estimated the statistical significance of Fourier spectral power peaks following \cite{1985Natur.317..416D,lou2003periodicities} and \citet{2018Periodic}.
The daily CME data were picked randomly within the time sequence to form an artificially randomized sequence with much reduced coherent periodic signals. In the general case of a normally distributed noise source, its power spectrum has an exponential distribution. Thus the natural logarithm of the cumulative distribution of the power decreases linearly with power. The slope of the logarithm of the cumulative distribution of power is a measure of the variance $\sigma ^{2}$ of the spectrum \citep{1985Natur.317..416D,lou2003periodicities,2018Periodic}. And then, we showed the natural logarithm of probability for spectral powers of both the actual daily CME counts and randomized sequence. The relevant $\sigma$ levels of fluctuation amplitudes were also shown. Comparing the two curves indicated that for powers greater than the 3$\sigma$ level, the CME data depart from random noise. To further assess the statistical significance of power peaks in this approach, we estimated the detection threshold in spectral power with a false alarm probability $p$$_{0}$. Finally, the Fourier spectra were tested for peaks at the 95\% confidence level and marked in the probability for spectral powers. We ran the same test procedure for the other type of CMEs and obtained qualitatively similar properties. 

2. We used the PyCWT\footnote{https://pycwt.readthedocs.io/en/latest/index.html}(version 0.3.0a22) for continuous wavelet spectral analysis. PyCWT is a Python module based on \citet{1998A}. We used the ``Morlet'' as the mother wavelet function and adopted $\omega _{0} = 6.0$ because the Fourier period and the wavelet time scale are nearly equal for this value \citep{farge1992wavelet,1998A}. The result of the wavelet analysis of a given dataset was visualized as a time-frequency spectrum (so-called the local wavelet spectrum). However, the global wavelet spectrum, as a frequency spectrum, can be directly compared to the Fourier spectrum. It has been suggested that the global wavelet spectrum could provide a useful measure of the background spectrum, against which peaks in the local wavelet spectrum could be tested \citep{kestin1998time}. 

We chose from the noise of the classic models as a background spectrum to detect the real periodicities from the power spectral peaks. Previous studies \citep{oliver1996rescaled,frick1997wavelet,1998A} remarked that the red noise model is more appropriate to physical phenomena. The best fit between the spectrum of the red noise model and that of the observed time series is obtained by taking the lag-1 auto-correlation coefficient of the latter for estimating a coefficient of correlation $r$. Eventually, the criterion of statistically significant signals at the 95\% level against the red noise background is used in both time-frequency spectra and global wavelet spectra.

The reliability of the results depends not only on the 95\% confidence level, but also on the cone of influence (COI). COI is the cross-hatched region in the local wavelet spectrum, where the edge effects on time series analysis cannot be ignored \citep{1998A,2004Application,2008Wavelet}. Therefore, the signals inside the COI must be considered virtual signals.

\section{Results}
\label{3}

\subsection{Periods for the Whole Time Intervals}
\label{3.1}

Figures \ref{fig:Figure1} and \ref{fig:Figure2} present the temporal evolution of the daily occurrence rate of the different types. It reveals that, in addition to following the $\approx$11-year Schwabe solar cycle, they also exhibit oscillations at higher frequencies. We searched for periodicities over the time intervals from January 1, 1996, to August 31, 2021, in 1) all CMEs, halo CMEs, partial halo CMEs, normal CMEs, and narrow CMEs (see Figure \ref{fig:Figure3}), and 2) classified them into the northern and southern hemispheres (see Figures \ref{fig:Figure4} and \ref{fig:Figure5}). In the following local wavelet spectra, wavelet power amplitudes are quantified by the color where the color ranges from purple (minimum power) to bright yellow (maximum power). Table \ref{tab:3} presents the results detected by the Fourier and global wavelet spectra (within COI) separately, which offers a practical overview of the results allowing easy qualitative comparison. The error bar for both two methods is determined as half of the full width at half-maximum (FWHM) about a power peak \citep{lou2003periodicities,2009A&A503L25W,2018A&A617A86L,2020ApJ88853L}. The error bars are taken into account in the following results and discussion section.

From the result of Fourier spectra (see Table \ref{tab:3}), periods of $\approx$6 months, $\approx$1.1 years, and $\approx$2.4 years are found in all CMEs and both hemispheres. Partial halo CMEs have periods of $\approx$1.3 years, 3.4 years, and 8.6 years in both hemispheres. Periods of $\approx$6 months, $\approx$1.1 years, $\approx$2.4 years, and 8.6 years are seen in both hemispheres for the normal CMEs. And only the peaks at 2.9 months, 11.4 months, $\approx$1.2 years, and $\approx$1.9 years in both hemispheres for the narrow CMEs.

As expected, the global wavelet spectra represent highly smoothed versions of the Fourier spectra. Indeed, the averaging process results in  merging of close peaks in broadly enhanced maxima. In some cases, these peaks exceed the red noise criterion, while the individual peaks do not. Only a few of the periods in the global wavelet spectrum that meet the red noise criterion are the same as those obtained from the Fourier spectrum (see Table \ref{tab:3}).

\subsection{Periods for SC23 and SC24}
\label{3.2}

In Figure \ref{fig:Figure6}, we split the wavelet spectra into two-time intervals corresponding to SC23 and SC24. In Figure \ref{fig:Figure7} and \ref{fig:Figure8}, the wavelet spectra show the northern and southern hemispheres corresponding to SC23 and SC24, thus allowing us to refine our analysis. These results are summarized in Table \ref{tab:4}. In terms of all CMEs, the period of 2.3-year is  presented in SC23, whereas a larger period of 3.7 years is presented in SC24. There are no identical periods for all CMEs in the northern hemisphere in both solar cycles. However, approximately 6.1 months and 2.2 years of all CMEs in the southern hemisphere can be found in both SC23 and SC24. In the case of the halo CMEs, only the periods of 10.3 months and 2.7 years are not confirmed for the two solar cycles. And in the case of the partial halo CMEs, there appear periods of 8.7 months and 2.9 years in SC24 which are absent in SC23. In the northern hemisphere, the partial halo CMEs have the same period of $\approx$6.5 months in both SC23 and SC24. While in the southern hemisphere, there are not the same period in both SC23 and SC24. Finally, turning our attention to the normal CMEs, we find more wealthy periodicities in SC23: 6.1 months, 1.1 years, and 2.3 years, while only the period of 5.8 months is observed in SC24. In the northern-normal CMEs, we detect three identical periods in both SC23 and SC24, i.e. 6.1 months, $\approx$2.4 years, and 3.7 years. In the southern-normal CMEs, we find only two, i.e. $\approx$6.1 months and 2.2 years. In the case of the narrow CMEs, several periods are significant: approximately 2.9 months, 6.5 months, 11 months, 2.2 years, and 3.9 years in SC24. However, higher frequencies of oscillations are less common in SC23. Regarding narrow CMEs in the northern hemisphere, we find the period of 3.9 years in both SC23 and SC24. And in the southern hemisphere, the peak at $\approx$2.5 years exits in both SC23 and SC24.

\begin{figure}
\begin{center}
	\includegraphics[width=\textwidth]{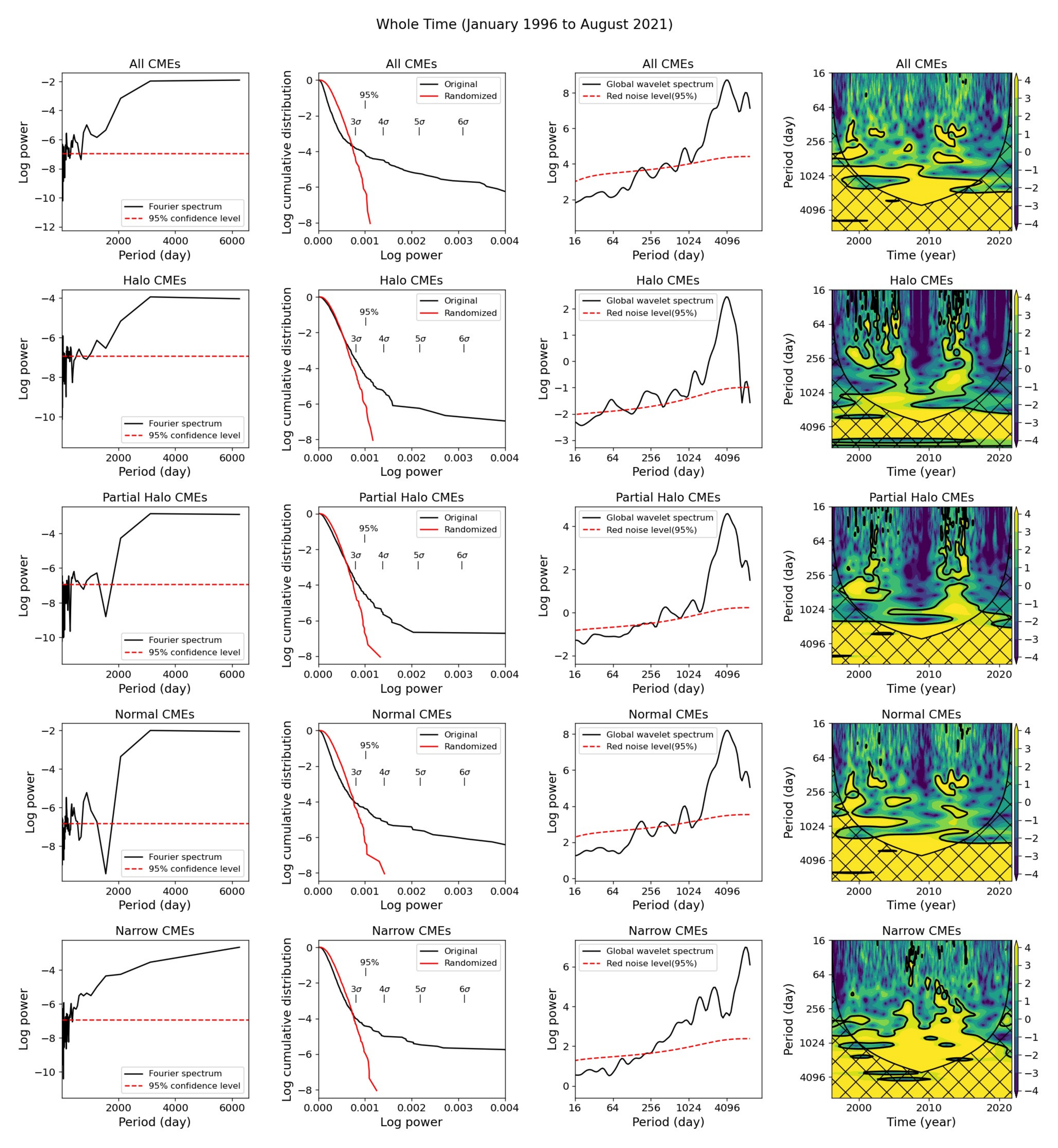}
\end{center}
    \caption{Fourier spectra (the first column), illustration of the method to estimate the significance of Fourier spectral power peaks (the second column), global wavelet spectra (the third column), and local wavelet spectra (the fourth column) for different angular widths of daily CMEs. In the Fourier spectra, the dashed red lines correspond to the 95\% confidence level. The second column displays the logarithm of the cumulative probability of the spectra versus power for the actual daily CME counts and the randomized sequence. In the global wavelet spectra, the dashed red lines correspond to the 95\% confidence level against the red noise background. Moreover, regions, where the local wavelet spectral power is statistically significant at the 95\% level against the red noise backgrounds, are contoured by black, thick lines. The first/second/third/fourth/fifth panels show all/halo/partial halo/normal/narrow CMEs, respectively.}
    \label{fig:Figure3} 
\end{figure}

\begin{figure}
\begin{center}
	\includegraphics[width=\textwidth]{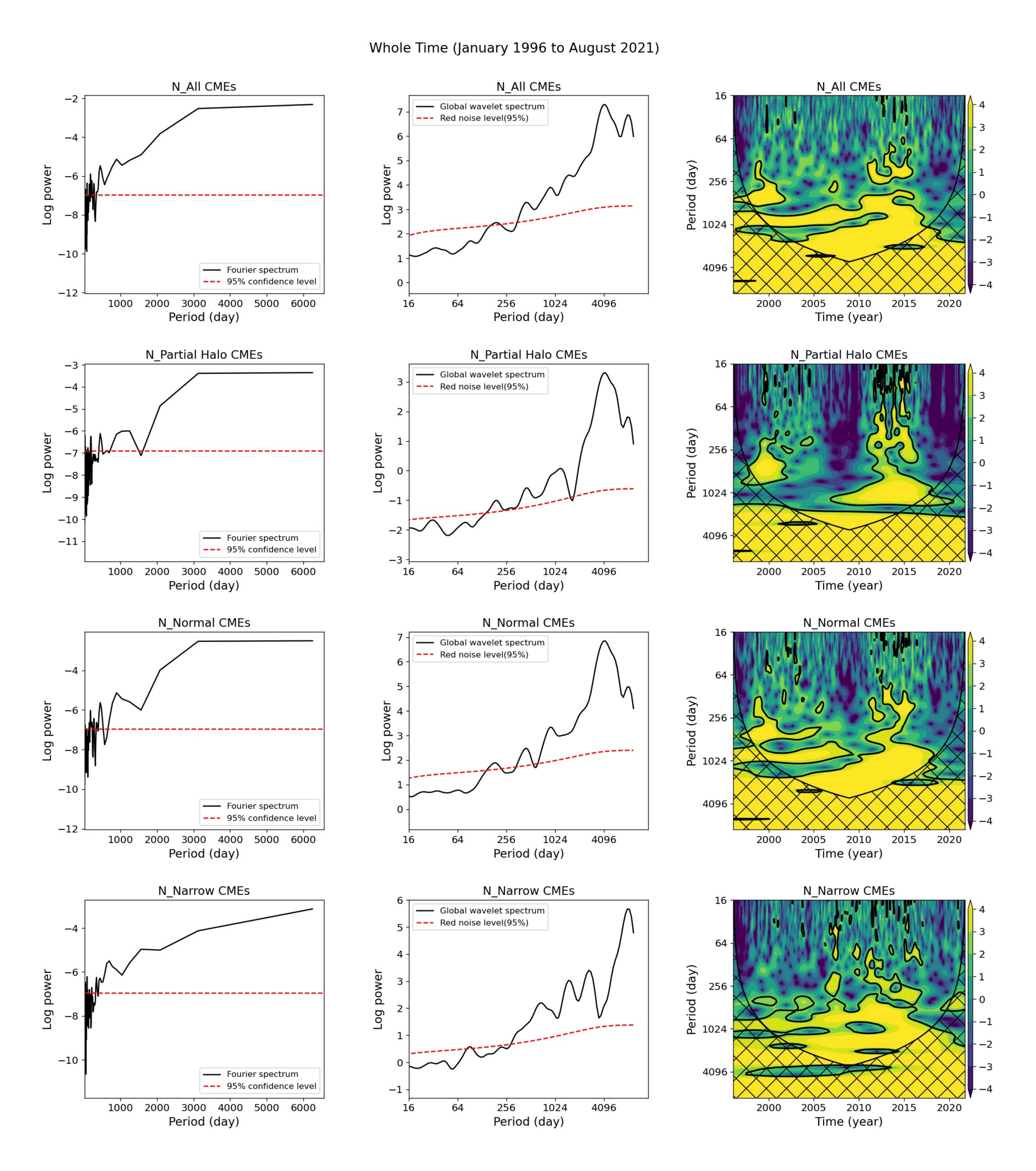}
\end{center}
    \caption{Fourier spectra (left column), global wavelet spectra (middle column), and local wavelet spectra (right column) for different angular widths of daily CMEs in the northern hemispheres. The first/second/third/fourth panels show all/partial halo/normal/narrow CMEs, respectively.}
    \label{fig:Figure4} 
\end{figure}

\begin{figure}
\begin{center}
	\includegraphics[width=\textwidth]{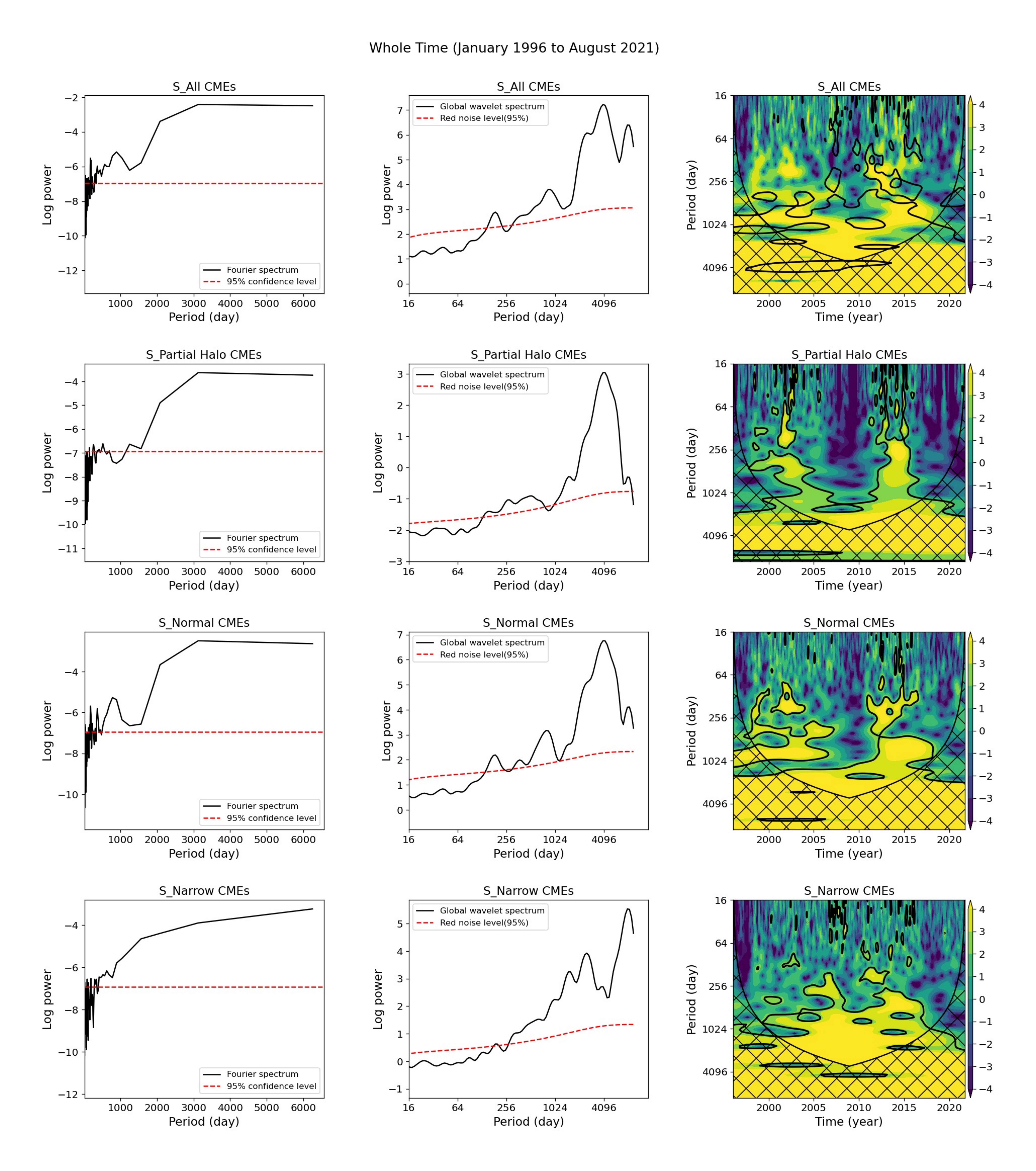}
\end{center}
    \caption{Fourier spectra (left column), global wavelet spectra (middle column), and local wavelet spectra (right column) for different angular widths of daily CMEs in the southern hemispheres. The first/second/third/fourth panels show all/partial halo/normal/narrow CMEs, respectively.}
    \label{fig:Figure5} 
\end{figure}

\begin{figure}
\begin{center}
	\includegraphics[width=0.9\textwidth]{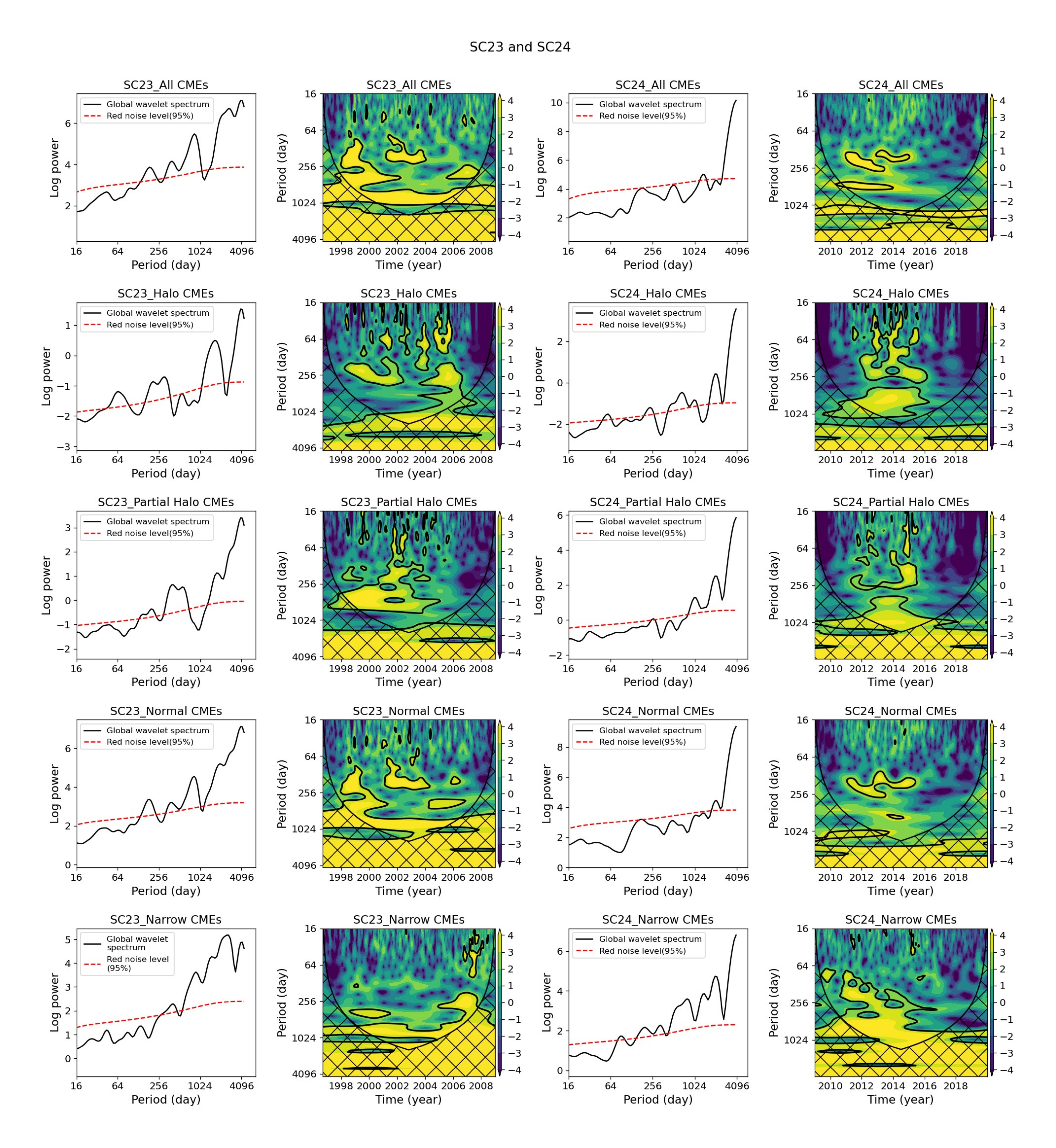}
\end{center}
    \caption{Global wavelet spectra and local wavelet spectra for different angular widths of daily CMEs for SC23 (the first and second columns) and SC24 (the third and fourth columns). The first/second/third/fourth/fifth panels show all/halo/partial halo/normal/narrow CMEs, respectively.}
    \label{fig:Figure6} 
\end{figure}

\begin{figure}
\begin{center}
	\includegraphics[width=\textwidth]{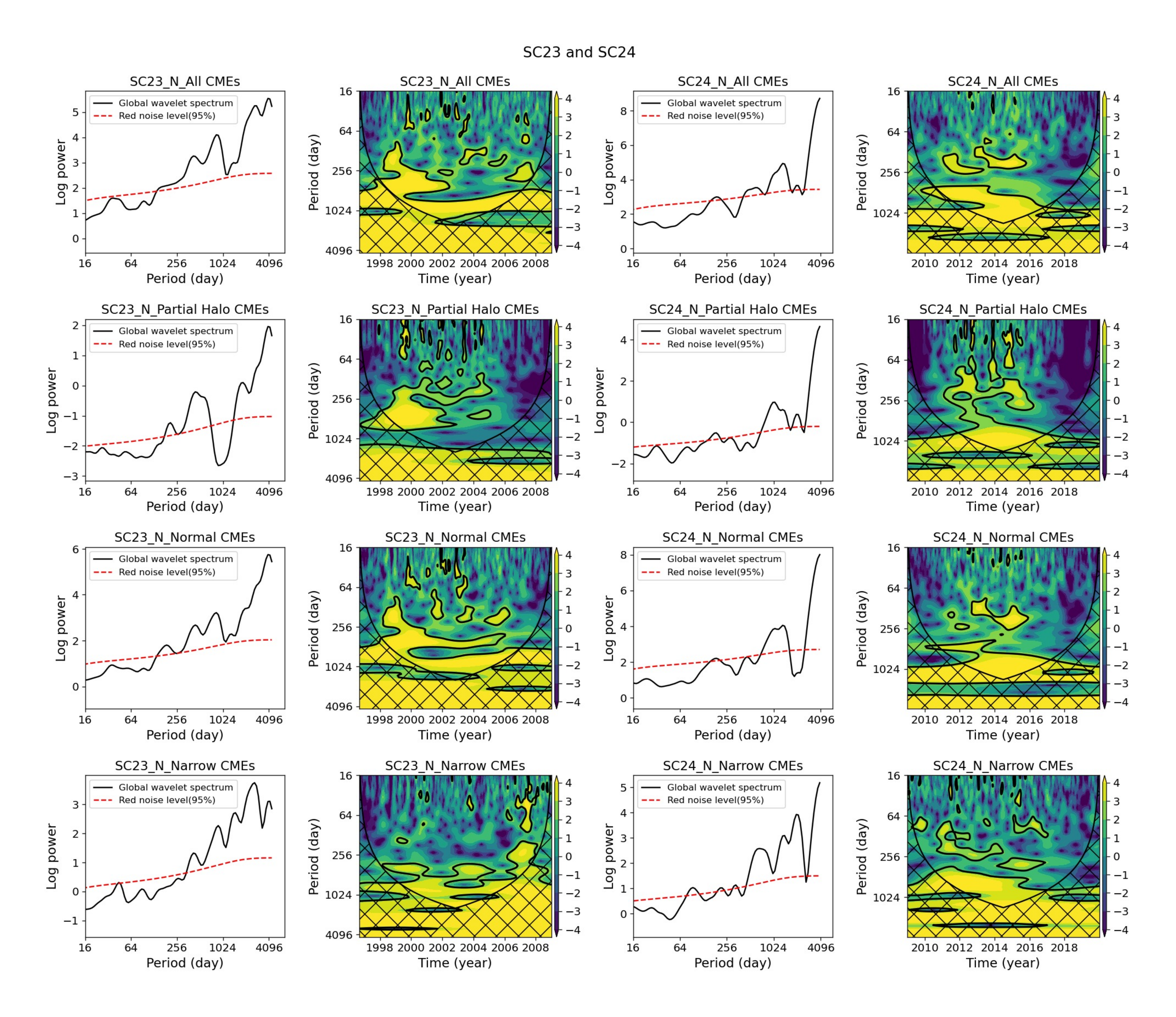}
\end{center}
    \caption{Global wavelet spectra and local wavelet spectra for different angular widths of daily northern CMEs for SC23 (the first and second columns) and SC24 (the third and fourth columns). The first/second/third/fourth panels show all/partial halo/normal/narrow CMEs, respectively.}
    \label{fig:Figure7} 
\end{figure}

\begin{figure}
\begin{center}
	\includegraphics[width=\textwidth]{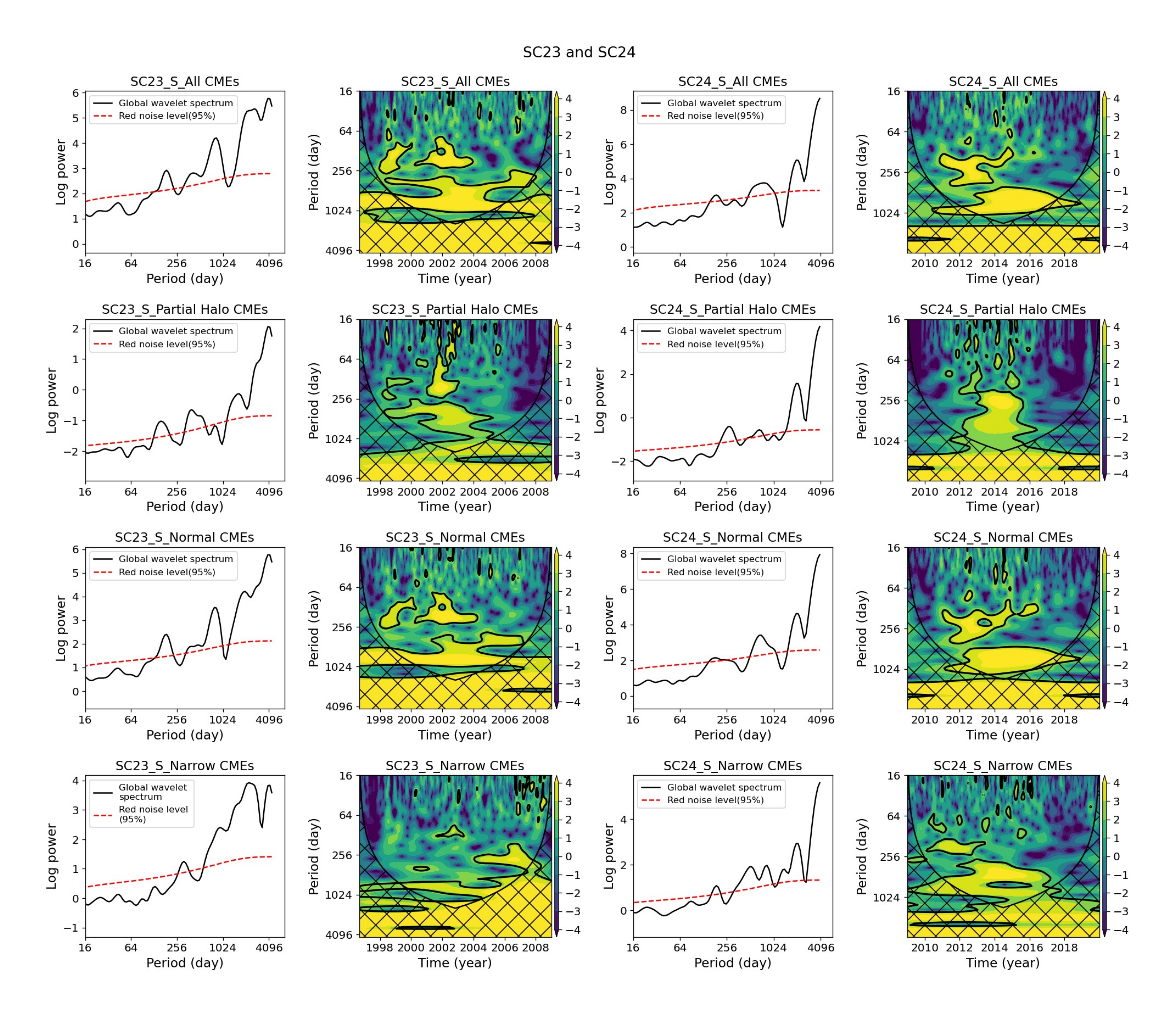}
\end{center}
    \caption{Global wavelet spectra and local wavelet spectra of different angular widths of daily southern CMEs for SC23 (the first and second columns) and SC24 (the third and fourth columns). The first/second/third/fourth panels show all/partial halo/normal/narrow CMEs, respectively.}
    \label{fig:Figure8} 
\end{figure}

The results from local wavelet spectra (Figures \ref{fig:Figure3}, \ref{fig:Figure4}, \ref{fig:Figure5}, \ref{fig:Figure6}, \ref{fig:Figure7} and \ref{fig:Figure8}) show more precisely the time intervals during which the various periods prevail. The characteristics of the detected periods are: i) during the ascending phase of the solar cycle and the vicinity of the cycle maxima: 0.7 months-2.1 years, ii) in the descending phase of the solar cycle: 2.2-2.9 years, iii) lasts from the descending phase of SC23 to the ascending phase of SC24: 3.1-8.6 years, and finally iv) the periods of $\approx$11 and 22 years which is significant but locates outside the COI curve.

\begin{table}
\centering
\renewcommand\tabcolsep{10pt} 
\caption{List of periods and the corresponding error bars over the time intervals (from January 1, 1996, to August 31, 2021). The confidence level of the respective Fourier and global wavelet spectral power peaks is 95\%. The results of the global wavelet spectrum are shown in parentheses, and the same period and error bars are shown in bold when detected by both methods.}
\label{tab:3}
\begin{tabular}{lccccc}
\hline
                & 1 Mo - 11 Mo               & 1 Yr - 1.4 Yr       & 1.5 Yr - 2.9 Yr     & 3 Yr - 5 Yr  & 6 Yr - 12 Yr \\
\hline
All             & 1.2±0.01 Mo,3.1±0.04 Mo   & \textbf{1.2±0.1 Yr} & 2.4±0.4 Yr          & -            & ($\approx$11 Yr)            \\
                & 6±0.2 Mo,(6.1±0.4 Mo)     &                     & (2.4±0.1 Yr)        &              &              \\
Halo            & 1.8±0.02 Mo,(2.1±0.1 Mo)  & -                   & 1.7±0.2 Yr          & 3.4±0.7 Yr   & \textbf{$\approx$11 Yr}   \\
                & 6±0.2 Mo,7.3±0.3 Mo       &                     & (1.7±0.1 Yr)        & (3.4±0.2 Yr) &              \\
                & (7.3±0.4 Mo),10.8±0.6 Mo  &                     &                     &              &              \\
Partial halo    & 1.2±0.01 Mo,6.4±0.2 Mo    & 1±0.1 Yr            & -                   & (3.3±0.2 Yr) & \textbf{$\approx$11 Yr}   \\
                & (6.5±0.4 Mo),(8.2±0.5 Mo) & 1.2±0.1 Yr          &                     & 3.4±0.7 Yr   &              \\
                & 8.5±0.4 Mo                & 1.4±0.1 Yr          &                     &              &              \\
Normal          & 1.2±0.01 Mo,5.9±0.2 Mo    & 1±0.1 Yr            & 2.4±0.4 Yr          & -            & \textbf{$\approx$11 Yr}   \\
                & (6.1±0.4 Mo)              & (1.1±0.1 Yr)        & (2.4±0.1 Yr)        &              &              \\
                &                           & 1.2±0.1 Yr          &                     &              &              \\
Narrow          & 2.9±0.04 Mo, 6.2±0.2 Mo   & 1.2±0.1 Yr          & (1.9±0.1 Yr)        & (4.1±0.2 Yr) & ($\approx$11 Yr) \\
                & (7.3±0.4 Mo),7.4±0.3 Mo   &                     & 2±0.2 Yr            &              &              \\
                & 11.4±0.6 Mo,(11.6±0.7 Mo) & \textbf{}           & \textbf{2.4±0.4 Yr} &              &              \\
\hline
All, N          & 2.9±0.1 Mo,6±0.2 Mo       & \textbf{1.2±0.1 Yr} & 2.4±0.4 Yr          & (4.1±0.2 Yr) & ($\approx$11 Yr)            \\
                & (6.5±0.4 Mo),7.1±0.2 Mo   &                     & (2.4±0.1 Yr)        &              &              \\
Partial halo, N & 1±0.004 Mo,6.4±0.2 Mo     & \textbf{1.2±0.1 Yr} & -                   & (3.3±0.2 Yr) & \textbf{$\approx$11 Yr}            \\
                & (6.5±0.4 Mo)              &                     &                     & 3.4±0.7 Yr   &              \\
Normal, N       & 1.1±0.01 Mo,6±0.2 Mo      & \textbf{1.2±0.1 Yr} & 2.4±0.4 Yr          & -            & \textbf{$\approx$11 Yr}            \\
                & (6.1±0.4 Mo),8.9±0.4 Mo   & \textbf{}           & (2.4±0.1 Yr)        &              &              \\
Narrow, N       & 1.4±0.01 Mo, 2.9±0.04 Mo  & 1.2±0.1 Yr          & (1.8±0.1 Yr )       & (4.1±0.2 Yr) & ($\approx$11 Yr) \\
                & (2.9±0.2 Mo), 4.9±0.1 Mo  &                     & 1.9±0.2 Yr          & 4.3±1.1 Yr   &              \\
                & 7.1±0.2 Mo, 11.4±0.6 Mo   &                     &                     &              &              \\
\hline
All, S          & 1±0.004 Mo,5.9±0.2 Mo     & 1±0.1 Yr            & 2.4±0.4 Yr          & -            & \textbf{$\approx$11 Yr} \\
                & (6.1±0.4 Mo),6.4±0.2 Mo   & \textbf{1.4±0.1 Yr} & (2.4±0.1 Yr)        &                 \\
Partial halo, S & 5±0.1 Mo,(5.2±0.3 Mo)     & \textbf{1.4±0.1 Yr} & 2.1±0.2 Yr          & 3.4±0.7 Yr   & \textbf{$\approx$11 Yr}   \\
                & 8.5±0.4 Mo,(9.2±0.5 Mo)   &                     &                     & (4.1±0.2 Yr) &              \\
Normal, S       & 1±0.01 Mo,5.9±0.2 Mo      & \textbf{1±0.1 Yr}   & 2.1±0.3 Yr          & -            & \textbf{$\approx$11 Yr}   \\
                & (6.1±0.4 Mo),7.1±0.2 Mo   &                     & (2.2±0.1 Yr)        &              &              \\
Narrow, S       & 0.7±0.003 Mo,2.9±0.04 Mo  & 1.1±0.1 Yr          & 1.7±0.2 Yr          & (4.1±0.2 Yr) & ($\approx$11 Yr) \\
                & 4.1±0.1 Mo,(6.1±0.4 Mo)   & 1.4±0.1 Yr          & (1.7±0.1 Yr)        &              &              \\
                & 6.4±0.2 Mo, 9.8±0.5 Mo    &                     & (2.7±0.2 Yr)        &              &              \\
                & 11.4±0.6 Mo               &                     &                     &              &                 \\
\hline
\end{tabular}
\end{table}

\begin{table}
\centering
\renewcommand\tabcolsep{10pt} 
\caption{List of periods and the corresponding error bars for two solar cycles (SC23 and SC24). The results of the global wavelet spectrum (within COI) are shown below with a confidence level of 95\% against the red noise backgrounds.}
\label{tab:4}
\begin{tabular}{lccccc}
\hline
                      & 1 Mo - 11 Mo           & 1 Yr - 1.4 Yr & 1.5 Yr - 2.9 Yr & 3 Yr - 5 Yr & 6 Yr - 12 Yr \\
\hline
All, SC23             & 6.1±0.4 Mo            & 1.1±0.1 Yr    & 2.3±0.1 Yr      & -           & -            \\
All, SC24             & 6.1±0.4 Mo            & 1.2±0.1 Yr    & -               & 3.7±0.2 Yr  & -            \\
Halo, SC23            & 2.2±0.1 Mo,6.9±0.4 Mo & -             & 1.5±0.2 Yr      & -           & -            \\
                      & 10.3±0.6 Mo           &               &                 &             &              \\
Halo, SC24            & 1.9±0.1 Mo,7.7±0.4 Mo & -             & 1.8±0.1 Yr      & -           & -            \\
                      &                       &               & 2.7±0.2 Yr      &             &              \\
Partial halo, SC23    & 4.9±0.3 Mo,6.5±0.4 Mo & 1.1±0.1 Yr    & 1.5±0.1 Yr      & -           & -            \\
Partial halo, SC24    & 8.7±0.5 Mo            & -             & 2.9±0.2 Yr      & -           & -            \\
Normal, SC23          & 6.1±0.4 Mo            & 1.1±0.1 Yr    & 2.3±0.1 Yr      & -           & -            \\
Normal, SC24          & 5.8±0.3 Mo            & -             & -               & -           & -            \\
Narrow, SC23          &                       & 1.2±0.1 Yr    & 2.4±0.1 Yr      & 4.1±0.2 Yr  & -            \\
Narrow, SC24          & 2.9±0.2 Mo,6.5±0.4 Mo & -             & 2.2±0.1 Yr      & 3.9±0.2 Yr  & -            \\
                      & 11±0.7 Mo             &               &                 &             &              \\
\hline
All, N, SC23          & -                     & 1.2±0.1 Yr    & 2.3±0.1 Yr      & -           & -            \\
All, N, SC24          & 6.1±0.4 Mo            & -             & 1.6±0.1 Yr      & 3.7±0.2 Yr  & -            \\
Partial halo, N, SC23 & 6.9±0.4 Mo            & 1.2±0.1 Yr    & -               & -           & -            \\
Partial halo, N, SC24 & 6.1±0.4 Mo            & -             & 2.9±0.2 Yr      & -           & -            \\
Normal, N, SC23       & 6.1±0.4 Mo            & 1.2±0.1 Yr    & 2.3±0.1 Yr      & 3.7±0.2 Yr  & -            \\
Normal, N, SC24       & 6.1±0.4 Mo            & -             & 2.4±0.1 Yr      & 3.7±0.2 Yr  & -            \\
Narrow, N, SC23       & 1.4±0.1 Mo            & 1.2±0.1 Yr    & 2.4±0.1 Yr      & 4±0.2 Yr    & -            \\
Narrow, N, SC24       & 3.1±0.2 Mo,7.3±0.4 Mo & -             & 1.8±0.1 Yr      & 3.9±0.2 Yr  & -            \\
                      & 11±0.7 Mo             &               &                 &             &              \\
\hline
All, S, SC23          & 6.1±0.4 Mo            & 1.2±0.1 Yr    & 2.3±0.1 Yr      & -           & -            \\
All, S, SC24          & 6.1±0.4 Mo            & -             & 2.2±0.1 Yr        & -           & -            \\
Partial halo, S, SC23 & 5.2±0.3 Mo            & 1.1±0.1 Yr    & -               & -           & -            \\
Partial halo, S, SC24 & 9.2±0.5 Mo            & -             & 1.8±0.1 Yr      & -           & -            \\
Normal, S, SC23       & 6.1±0.4 Mo            & 1.1±0.1 Yr    & 2.2±0.1 Yr      & -           & -            \\
                      &                       & 1.3±0.1 Yr    &                 &             &              \\
Normal, S, SC24       & 5.8±0.3 Mo,7.2±0.5 Mo & -             & 2.2±0.1 Yr      & -           & -            \\
Narrow, S, SC23       & 10.3±0.3 Mo           & -             & 2.7±0.2 Yr      & -           & -            \\
Narrow, S, SC24       & 6.1±0.4 Mo            & 1.4±0.1 Yr    & 2.3±0.2 Yr      & 3.9±0.2 Yr  & -            \\
\hline
\end{tabular}
\end{table}

\section{DISCUSSIONS}
\label{4}
Quasi-periodic results for different categories of CMEs reveal an extremely complex situation. There are similarities and differences in these periods. Some periods may reveal underlying physical causes. We will discuss qualitative, reasonable physical connections.

Based on the results in Table \ref{tab:3} and \ref{tab:4}, we combined periods from the Fourier and global wavelet spectra in the following comparison. We find some commonalities and differences for less than one year. For example, the solar rotation period of 0.8-3.3 months is found in all types except for the southern-partial halo CMEs. The 2 or 3.3-month period of different types CMEs might be due to the sampling rate of the data series or an artifact of projection effects. Even if a CME-productive active region produces CMEs randomly, we might still get a 2 or 3.3-month period since halo or partial halo CMEs happen when the active region is around the solar disk center. The period of solar rotation is about 27 days (roughly one month). Different types of CMEs might therefore possess a period of 0.8-3.3 months. 

Our results show that the Rieger-type periodicity (corresponding to the 4.1-6.5 months periods we obtained) is observed in all types. The Rieger periodicity is also found in high-energy solar flares \citep{RN85Rieger}. Similar periodicities have been detected in the photospheric magnetic flux during cycles 13, 15, 17, 18, and 20 \citep{ballester2002near}. \citet{2006A&A...452..647J} found the Rieger periodicity in the sunspot number and area in the northern hemisphere during SC23. \citet{carbonell1992periodic} proposed a scenario in which the quasi-periodic emergence of magnetic flux can trigger flares (via magnetic reconnection). This scenario could be directly applied to CMEs. And it has been suggested that the Rieger periodicity is due to equatorially trapped Rossby waves \citep{lou2000rossby}. In any case, the Rieger periodicity found in the different CME types confirms the existence of a global process that gives rise to magnetic flux escape from subphotospheric regions, through the solar atmosphere, to the heliosphere. It is easy to understand that CMEs are related to the processes of accumulation and dissipation of magnetic free energy from the interior of the Sun to the solar atmospheric layers and the interplanetary space \citep{2019ApJ88051M}.

The period of $\approx$9 months (10 rotations) can only be found in the partial halo CMEs, southern-partial halo CMEs, northern-normal CMEs, and southern-narrow CMEs. However, we only identify this periodicity in both partial halo CMEs and its southern hemisphere in SC24. \citet{getko2008mid} found that the sunspot area fluctuations in the northern and the southern hemispheres show quasi-periodicity in 9 months. Moreover, the result of \citet{getko2014ten} reported that this period is mainly detected during the high-activity periods, which is the same as our results of local wavelet spectra. It may reveal the special characteristics of partial halo CMEs compared to other angular widths. Moreover, we suggest that partial halo CMEs are more correlated with sunspot activity. Partial halo CMEs represent CMEs with higher kinetic energy (generally speaking, the higher the velocity of the CME, the higher the kinetic energy, and the wider the outward extension). Partial halo CMEs are more closely associated with active regions. Especially those with stronger magnetic fields, they tend to produce wider CMEs. Therefore, partial halo CMEs have the potential to produce the same 9-month period as sunspots.

The periods in the range 1-2.9 years seem to be present in each type. In particular, during SC23, the sunspot number (SSN), area and the magnetic flux (open and total) show such behavior \citep{kane2005short}. And these periodicities are also detected from interplanetary magnetic field and geomagnetic Ap index \citep{tsichla2019spectral}. Moreover, it was shown \citep{mursula2003mid,knaack2005spherical,knaack2005evolution} that oscillations with a period of around 1.3 years are a single process that manifests itself at all levels from the tachocline and photosphere (e.g., areas and numbers of sunspots, large-scale magnetic fields) up to the Earth’s magnetosphere (geomagnetic activity) and the far heliosphere (cosmic rays). The period of 1.7 years is identified in halo CMEs, narrow CMEs, northern-narrow CMEs, and southern-narrow CMEs. However, if the data are analyzed in each solar cycle, a similar period is only found in halo CMEs in both SC23 and SC24. Furthermore, this period was reported to be present in interplanetary plasma parameters and cosmic ray intensity as a modulation effect of solar activity \citep{kudela2002time,kato20031}. It can be found that the two periods, 1.3 years and 1.7 years, are mainly focused on the parameters of various types of solar radiation and plasma in an interplanetary environment, which are usually considered to be related to the energy released from the Sun. Therefore, it may suggest the important role of halo CMEs in the interplanetary environment and space weather \citep{2015ApJ804L23G,2020ApJ903118D}. 

Usually, for the QBO periodicities (around two years) are associated with the following probable mechanisms: (i) the quasi-two-year impulse of shear waves could bring out the quasi-two-year periodicity of the photospheric differential rotation \citep{pataraya1995generation}, but this mechanism can work only around the solar minima; (ii) the presence of two types of dynamo actions operating at different depths, one near the top of the layer extending from the surface down to 5\% below it and the other one seated at the base of the convection zone \citep{benevolenskaya1998model}; (iii) the spatio-temporal fragmentation of the nonlinear dynamo processes in the convection zone \citep{covas2000spatiotemporal}; (iv) the beating between a dipole and quadrupole magnetic configuration of the dynamo \citep{simoniello2013quasi}; and (v) magnetic Rossby waves in the solar tachocline \citep{lou2003periodicities,zaqarashvili2010quasi,mcintosh2015solar}.

For normal CMEs and both hemispheres, periods around 3.3-4.3 years are relatively uncommon. We can only find these periods in the northern hemispheres of both SC23 and SC24. Besides, we obtain $\approx$11 years sunspot cycle (Schwabe cycle), which is prevalent in all types.

 These periodic results show the characteristics of CMEs of different angular widths: various periodicity, intermittency, asymmetric development in the two hemispheres. For CMEs of different angular widths, there are indeed various periods: 9 months, 1.7 years, and 3.3-4.3 years. We suggest that the behaviors of the periodic differences may reflect the discrepancies in CMEs with different angular widths. \citet{2005AcASn46416S} showed the relationship between CMEs and solar microwave bursts (SMBs) and found that the SMBs associated with normal/narrow CMEs have relatively short durations. While those correlated with halo/partial halo CMEs have either short or long durations probably, which may indicate effects of different magnetic field structures and magnetic energy storages on the solar bursts. \citet{yashiro2003properties} studied the statistical properties of narrow CMEs and reported that they do not constitute a subset of normal CMEs and have different acceleration mechanisms. Moreover, morphologically, narrow CMEs show an elongated jet-like shape. In contrast, normal CMEs present a closed (or convex-outward) loop \citep{Chen2011Coronal}.

In past studies of \citet{lou2003periodicities}, \citet{2008Short}, \citet{2014Different}, \citet{2014Wavelet}, \citet{2018Periodic} and \citet{lamy2019coronal}, among a dozen occurrence rate of all CMEs revealed by spectral analyses, we confirm the periods of 1.2(±0.01) months, 3.1(±0.04) months, $\approx$6.1(±0.4) months, 1.2(±0.1) years, and 2.4(±0.4) years. We do find additional periods of all CMEs only in one hemisphere or during a specific solar cycle. For example, 7.1(±0.2) months and 4.1(±0.2) years in the northern hemisphere, 1(±0.004) months, 5.9(±0.2) months, 1(±0.1) years, 1.4(±0.1) years, and 2.4(±0.4) years in the southern hemisphere, 6.1(±0.4) months in SC23 and 6.1(±0.4) months, 1.2(±0.1) years, and 3.7(±0.2) years in SC24. It is probably partly due to the fact that the period search is inherently difficult, which is why many different methods have been developed, and a safe approach is to implement several methods to determine the robustness of the results. In the case of non-stationary processes, it is even more complicated, and the assumed periodic patterns might change or even disappear over time, which is certainly the case with solar phenomena, as this becomes more and more evident as observations accumulate (as the results of the periods in Table \ref{tab:1}). Alternatively, it can be due to the fact that the CDAW catalog we use (see Figure 1 in \citet{2014Different}) itself is to some extent inconsistent with other catalogs (see Figure 8 in \citet{2014JGRA11947L}). Most importantly, we use the CME angular width to classify the CMEs.

The angular width of a CME is an important parameter in considering its space weather impact. This is reflected in the fact that in addition to its contribution to whether and when CMEs reach the Earth and how strong the geomagnetic disturbances introduced are, as well as determining the spatial extent of the impact in front of CMEs. The obtained information on the periods of different angular widths can be a good reference to determine the impact on the Earth in the spatial background. Besides, the speed of a CME is considered to be a more prominent parameter in terms of the effects on space weather \citep{kilcik2020temporal}. In future studies, periodic analysis of speeds with different angular widths can also be implemented.

\section{CONCLUSIONS}
\label{5}
In this work, we used data on the number of daily CMEs from January 1, 1996, to August 31, 2021. The CMEs were classified by angular width. Frequency and time-frequency analysis methods were used to systematically analyze the quasi-periodic variations corresponding to CMEs with different angular widths in the northern and southern hemispheres. We summarized below the main findings obtained from our analysis.
\begin{enumerate}[(1)]
    \item For CMEs of different angular widths, there are indeed various periods: 9 months, 1.7 years, and 3.3-4.3 years.
	\item Compared with previous studies using the occurrence rate of CMEs, we obtain the same periods of 1.2(±0.01) months, 3.1(±0.04) months, $\approx$6.1(±0.4) months, 1.2(±0.1) years, and 2.4(±0.4) years.
	\item After classifying CMEs by angular width, we do find additional periods of all CMEs only in one hemisphere or during a specific solar cycle. For example, 7.1(±0.2) months and 4.1(±0.2) years in the northern hemisphere, 1(±0.004) months, 5.9(±0.2) months, 1(±0.1) years, 1.4(±0.1) years, and 2.4(±0.4) years in the southern hemisphere, 6.1(±0.4) months in SC23 and 6.1(±0.4) months, 1.2(±0.1) years, and 3.7(±0.2) years in SC24.
	\item The results of our quasi-periodic analysis show that the occurrence rate of CMEs exhibits statistically significant short- and medium-ranges oscillations characterized by various periodicity, intermittency, and asymmetric development in the northern and southern solar hemispheres.
\end{enumerate}

It is well known that solar magnetic fields are transported from the internal convection zone towards the surface layers of the Sun. Some of the magnetic fields is dissipated in the lower atmospheric layers (photosphere and chromosphere), and parts of them escape from the Sun toward the outer heliosphere in the form of CMEs and solar wind. The quasi-periodicities of the CMEs obtained by us were previously found in the solar flare activity \citep{kilcik2019comparison,kilcik2020temporal}, in the solar wind \citep{2000AdSpR..25.1939M}, in the interplanetary magnetic field \citep{2004SoPh..221..337M}, and in the geomagnetic activity \citep{mursula2003mid,2017JGRA..122.5043O}. Therefore, the quasi-periodic variations of the CMEs should be a connecting agent among the oscillations in the coronal magnetic activity, the solar flare eruption, and interplanetary space.

\begin{acknowledgments}

This work is supported by the National SKA Program of China No 2020SKA0110300, the Joint Research Fund in Astronomy (U1831204, U1931141) under cooperative agreement between the National Natural Science Foundation of China (NSFC) and the Chinese Academy of Sciences (CAS), the National Science Foundation for Young Scholars (11903009). Funds for International Cooperation and Exchange of the National Natural Science Foundation of China (11961141001). Fundamental and Application Research Project of Guangzhou(202102020677). The National Natural Science Foundation of China (No. 11873089), the CAS “Light in Western China”Program, and the Yunnan Province XingDian Talent Support Program. This work is also supported by Astronomical Big Data Joint Research Center, co-founded by National Astronomical Observatories, Chinese Academy of Sciences and Alibaba Cloud. 

We thank the anonymous reviewers for the careful reading, helpful comments, and constructive suggestions.

\end{acknowledgments}

\bibliography{referrence}



\end{document}